\documentstyle[12pt]{article}

\let\ni=\noindent

\begin{document}
\begin{flushright}
IFT-UW/97-15
\end{flushright}

\baselineskip 0.75cm
 
\pagestyle {empty}

\renewcommand{\thefootnote}{\fnsymbol{footnote}}

~~~

\vspace{1.0cm}

{\large \centerline {\bf Texture dynamics for neutrinos\footnote{Work supported 
in part by the Polish KBN--Grant 2--B302--143--06.}}}

\vspace{1.0cm}

{\centerline {\sc Wojciech Kr\'{o}likowski}}

\vspace{1.0cm}

{\centerline {\it Institute of Theoretical Physics, Warsaw University}}

{\centerline {\it Ho\.{z}a 69,~~PL--00--681 Warszawa, ~Poland}}

\vspace{1.0cm}

{\centerline {\bf Abstract}}

\vspace{0.3cm}

 An ansatz for mass matrix was recently proposed for charged leptons, predict%
ing (in its diagonal approximation) $ m_\tau \simeq 1776.80 $ MeV from the
experimental values of $ m_e $ and $ m_\mu $, in agreement with $ m_\tau^{\rm 
exp} = 1777.00^{+0.30}_{-0.27} $ MeV. Now it is applied to neutrinos. If the 
amplitude of neutrino oscillations $\nu_\mu \rightarrow \nu_\tau $ is $\sim 1/2
$ and $|m^2_{\nu_\tau} -m^2_{\nu_\mu}| \sim (0.0003\;{\rm to}\;0.01)\;\,{\rm 
eV}^2$, as seems to follow from atmospheric--neutrino experiments, this ansatz 
predicts $ m_{\nu_e} \ll m_{\nu_\mu} \sim (0.2\;\,{\rm to}\;\,1)\times 10^{-2}
$ eV and $ m_{\nu_\tau} \sim(0.2\;\,{\rm to} \;\,1)\times 10^{-1}\;\,{\rm eV}$,
and also the amplitude of neutrino oscillations $\nu_e \rightarrow \nu_\mu 
\sim 2^{+4}_{-2}\times 10^{-4}$ (in the vacuum). Such a very small amplitude 
for $\nu_e \rightarrow \nu_\mu $ is implied by the value of $ m_\tau^{\rm exp}
- 1776.80 $ MeV used to determine the deviation of the diagonalizing matrix 
$\widehat{U}^{(e)}$ from $\widehat{1}$ in the lepton Cabibbo---Kobayashi---%
Maskawa matrix $\widehat{V} = \widehat{U}^{(\nu)\,\dagger}\widehat{U}^{(e)}$. 
Here, $\widehat{U}^{(\nu)}$ by itself gives practically no oscillations $\nu_e 
\rightarrow \nu_\mu $, while it provides the large oscillations $\nu_\mu 
\rightarrow \nu_\tau $ .

\vspace{0.3cm} 

\ni PACS numbers: 12.15.Ff , 12.90.+b 
 
\vspace{3.0cm} 

\ni August 1997

\vfill\eject

\pagestyle {plain}

\setcounter{page}{1}

\vspace{0.2cm}

\ni {\bf 1. Introduction}

\vspace{0.3cm}

 First, let us say a few introductory words about two familiar notions of neut%
rino weak--interaction states and neutrino mass states.

 Since, apparently, neutrinos display no electromagnetic nor strong inter\-%
actions, experimental detectors select their weak--interaction states, what is 
in contrast to mass states selected by detectors in the case of charged leptons
and hadrons (built up from quarks).

 Thus, if the neutrino mass matrix $\widehat{M}^{(\nu)}$ and/or charged--lepton
mass matrix $\widehat{M}^{(e)}$ are originally nondiagonal in the bases

\vspace{0.15cm}

\begin{equation}
\vec{\nu}\,^{(0)} = \left(\begin{array}{c} {\nu}^{(0)}_e \\ {\nu}^{(0)}_\mu
\\ {\nu}^{(0)}_\tau \end{array} \right)
\end{equation}

\vspace{0.15cm}

\ni and

\vspace{0.15cm}

\begin{equation}
\vec{e}\,^{(0)} = \left(\begin{array}{c}  {e}^{\,-\,(0)} \\ {\mu}^{\,-\,(0)} \\  
{\tau}^{\,-\,(0)} \end{array} \right)\;\;,
\end{equation}

\vspace{0.15cm}

\ni respectively, the neutrino weak--interaction states

\vspace{0.15cm}

\begin{equation}
\vec{\nu}\, = \left(\begin{array}{c} {\nu}_e \\ {\nu}_\mu
\\ {\nu}_\tau \end{array} \right)
\end{equation}

\vspace{0.15cm}

\ni are, {\it mutatis mutandis}, analogues of the Cabibbo---Kobayashi---Maskawa
transforms

\vspace{0.15cm}

\begin{equation}
\vec{d}' = \left(\begin{array}{c} d' \\ s' \\ b' \end{array} \right)
\end{equation}

\vspace{0.15cm}

\ni of down--quark mass states

\begin{equation}
\vec{d}^{(m)} \equiv \vec{d} = \left(\begin{array}{c} d \\ s \\ b \end{array} 
\right)\; .
\end{equation}

\vspace{0.15cm}

\ni It is so, though $\vec{\nu}$ are experimentally observed states, in con%
trast to $\vec{d}' \neq \vec{d}$, where $\vec{d}^{(m)} \equiv \vec{d}$ describe
experimentally observed states.

 In fact, neutrino weak--interaction states are defined as

\vspace{0.15cm}

\begin{equation}
\vec{\nu} = \widehat{V}^{\,-1}\vec{\nu}\,^{(m)}\;\;,\;\; \widehat{V} = 
\widehat{U}^{(\nu)\,-1} \widehat{U}^{(e)}\;\, ,
\end{equation}

\vspace{0.15cm}

\ni where

\vspace{0.15cm}

\begin{equation}
\vec{\nu}\,^{(m)} = \widehat{U}^{(\nu)\,-1}\vec{\nu}^{(0)}\;\;,\;\;  
\widehat{U}^{(\nu)\,-1} \widehat{M}^{(\nu)} \widehat{U}^{(\nu)} = 
{\rm diag}\left(m_{\nu_e}\,,\;m_{\nu_\mu} \,,\;m_{\nu_\tau}\right)
\end{equation}

\vspace{0.15cm}

\ni are neutrino mass states

\vspace{0.15cm}

\begin{equation}
\vec{\nu}\,^{(m)} = \left(\begin{array}{c} {\nu}^{(m)}_e \\ {\nu}^{(m)}_\mu
\\ {\nu}^{(m)}_\tau \end{array} \right)\;\, ,
\end{equation}

\vspace{0.15cm}

\ni while

\vspace{0.15cm}

\begin{equation}
\vec{e}\,^{(m)} = \widehat{U}^{(e)\,-1}\vec{e}\,^{(0)}\;\;,\;\;\widehat{U}^{(e)\,
-1} \widehat{M}^{(e)} \widehat{U}^{(e)} = {\rm diag}\left(m_{e}\,,\;m_{\mu} \,,
\;m_{\tau}\right)
\end{equation}

\vspace{0.15cm}

\ni represent charged--lepton mass states

\begin{eqnarray}
\vec{e}\,^{(m)} \equiv \vec{e} = \left(\begin{array}{c} e^- \\ {\mu}^-
\\ {\tau}^- \end{array} \right)\;\, .
\end{eqnarray}

\ni Here, $\vec{e}\,^{(m)} \equiv \vec{e}$ describe experimentally observed 
states, in contrast to $ \vec{\nu}\,^{(m)} \neq \vec{\nu} $. It can be readily 
seen that the states $\vec{\nu}\,^{(m)}(t)$, as given in Eq. (7), are eigenstates
of the neutrino mass operator

\begin{equation}
\int d^3\vec{r}\sum_{i\,j} \nu^{(0)\,\dagger}_i(x) M_{i\,j}^{(\nu)}\nu_j^{(0)}
(x) = \int d^3\vec{r}\sum_{i}m_{\nu_i}\nu^{(m)\,\dagger}_i(x)\nu^{(m)}_i(x)\;,
\end{equation}

\ni where $ \widehat{M}^{(\nu)} = \left( M_{i\,j}^{(\nu)}\right) $. Similarly,
the states $\vec{e}\,^{(m)}(t) $, as defined in Eq. (9), represent eigenstates of
the charged--lepton mass operator. The unitary matrix $\widehat{V}$, introduced
in Eq. (6), is obviously a lepton analogue of the Cabibbo---Kobayashi---Maskawa
matrix, because the lepton charge--changing weak current has the form

\begin{equation}
\vec{\nu}\,^{(0)\,\dagger}(x)\beta\gamma^\mu(1\! -\!\gamma^5)\vec{e}\,^{(0)}(x)
\!=\!\vec{\nu}\,^{(m)\,\dagger}(x)\widehat{V}\beta\gamma^\mu(1\! -\!\gamma^5)
\vec{e}\,^{(m)}(x) = \vec{\nu}^{\,\dagger}(x)\beta\gamma^\mu(1\! -\!
\gamma^5)\vec{e}\,(x)\,,
\end{equation}

\ni where Eqs. (7), (9) and (6) are used.

 Note that the formula

\begin{equation}
\vec{\nu} = \widehat{U}^{(e)\,-1} \vec{\nu}\,^{(0)}
\end{equation}

\ni follows generally from Eqs. (6) and (7). This implies in the case when $
\widehat{M}^{(e)} $ is diagonal ({\it i.e.}, $\widehat{U}^{(e)} = \widehat{\bf 
1} $ and so, $\vec{e} = \vec{e}\,^{(0)}$) that $\vec{\nu} = \vec{\nu}\,^{(0)}$.
In this case, $\widehat{V} = \widehat{U}^{(\nu)\,-1}$ and thus $\vec{\nu} = 
\widehat{U}^{(\nu)} \vec{\nu}\,^{(m)}$, what means that $\vec{\nu} \neq \vec{
\nu}\,^{(m)}$ if $\widehat{U}^{(\nu)} \neq \widehat{1}$. When, alternatively, $
\widehat{M}^{(\nu)}$ is diagonal ({\it i.e.}, $\widehat{U}^{(\nu)} = \widehat{
\bf 1}$ and so, $\vec{\nu}\,^{(m)} = \vec{\nu}\,^{(0)}$), then Eq. (13) shows 
that $\vec{\nu}\, = \widehat{U}^{(e)\,-1} \vec{\nu}\,^{(m)}$ giving $\vec{\nu} 
\neq \vec{\nu}\,^{(m)}$ if $ \widehat{U}^{(e)}\neq \widehat{1} $.

 As is well known, neutrino mixing {\it i.e.}, the mixing of neutrino mass sta%
tes $\nu_i^{(m)}$ within neutrino weak--interaction states $\nu_i $, expressed
by the formula (6),

\begin{equation}
\nu_i = \sum_j\left(\widehat{V}^{\,-1} \right)_{i\,j} \nu_j^{(m)}\; ,
\end{equation}

\ni implies neutrino oscillations (in time) between states $\nu_i$. They occur 
if masses $m_{\nu_i}$ are not all degenerate and, of course, the mass matrices 
$\widehat{M}^{(\nu)}$ and/or $\widehat{M}^{(e)}$ are nondiagonal. In fact, 
since time--dependent weak--interaction neutrino states are

\begin{equation}
\nu_i(t) = e^{\,-iHt}\nu_i  = \sum_j\left(\widehat{V}^{\,-1} \right)_{i\,j} 
\nu_j^{(m)}e^{\,-iE_jt}\;,
\end{equation}

\ni the probability of oscillations $\nu_i \rightarrow \nu_j $ (in the vacuum)
is given by the formula

\begin{equation}
P(\nu_i \rightarrow \nu_j,t) = |\langle\nu_j |\nu_i(t)\rangle|^2 = \sum_{k\,l}
V^*_{j\,l}V_{i\,l}V_{j\,k}V^*_{i\,k} \exp\left(i\frac{m^2_{\nu_l}-m^2_{\nu_k}}
{2|\vec{p}|}\,t\right)\;,
\end{equation}

\ni where the ultrarelativistic relation 

\begin{equation}
E_l - E_k = \frac{m^2_{\nu_l}-m^2_{\nu_k}}{2|\vec{p}|}
\end{equation}

\ni is used for neutrino mass states. In Eq. (16), usually $ t/|\vec{p}| = L/E 
$, what is replaced by $ \,4\times 1.26693\,L/E\, $ if $m^2_{\nu_l}-m^2_{\nu_k}$, 
$ L $ and $ E $ are measured in eV$ ^2 $, km and GeV, respectively. Here, $ L $
is the source--detector distance.

 Concluding this introductory Section, we can see that the masses $ m_{\nu_e}\,
,\;m_{\nu_\mu}\,,\;m_{\nu_\tau}$ of neutrino mass states $\nu_e^{(m)}\,,\;
\nu_\mu^{(m)}\,,\;\nu_\tau^{(m)} $ as well as their mixing parameters [involved
in the lepton Cabibbo---Kobayashi---Maskawa matrix $\widehat{V} \equiv \left(
V_{ij}\right)$] can be determined, if neutrino and charged--lepton mass 
matrices $\widehat{M}^{(\nu)} \equiv \left(M^{(\nu)}_{ij}\right)$ and $\widehat
{M}^{(e)} \equiv \left(M^{(e)}_{ij}\right)$ are given explicitly. Once the 
mixing of $\nu_e^{(m)}\,,\;\nu_\mu^{(m)}\,,\;\nu_\tau^{(m)}$ (in weak inter%
actions) and the masses $ m_{\nu_e}\,,\;m_{\nu_\mu}\,,\;m_{\nu_\tau}$ are 
known, the oscillations (in time) between the neutrino weak--interaction states
$\nu_e\,,\;\nu_\mu\,,\;\nu_\tau$ can be evaluated.

 The notation $\nu_e^{(m)}\,,\;\nu_\mu^{(m)}\,,\;\nu_\tau^{(m)}$ for neutrino 
mass states and $ m_{\nu_e}\,,\;m_{\nu_\mu}\,,\;m_{\nu_\tau}$ for their masses,
though consequent, may be sometimes confusing about its difference with $\nu_e
\,,\;\nu_\mu\,,\;\nu_\tau $ being the neutrino weak--interaction states to 
which masses cannot be ascribed. Thus, in the case of mass states, the notation
$\nu_0\,,\;\nu_1\,,\;\nu_2 $ and $ m_{\nu_0}\,,\;m_{\nu_1}\,,\;m_{\nu_2}$ is, 
perhaps, more adequate. We hope, however, that the Reader will not be seriously
confused by the former notation used consequently throughout this paper (notice
that in the Particle Tables of Ref. [2] the neutrino masses are also denoted by 
$ m_{\nu_e}\,,\;m_{\nu_\mu}\,,\;m_{\nu_\tau}$).

 In the next Section, an ansatz for the mass matrices $\widehat{M}^{(e)}$ and
$\widehat{M}^{(\nu)}$ will be described and its cosequences derived. This 
ansatz introduces a kind of "texture dynamics" for leptons.

\vspace{0.3cm}

\ni {\bf 2. A model for $\widehat{M}^{(e)}$ and $\widehat{M}^{(\nu)}$}

\vspace{0.3cm}

 Let us consider the following ansatz [1] for charged--lepton and neutrino mass
matrices:

\begin{equation}
\widehat{M}^{(e,\nu)} = \widehat{\rho}\,\widehat{h}^{(e,\nu)}\,\widehat{\rho}
\end{equation}

\ni with

\begin{eqnarray}
\widehat{h}^{(e,\nu)} & = & \mu^{(e,\nu)}\left[\widehat{N}^{2} - (1 - 
\varepsilon^{(e,\nu)\,2})\widehat{N}^{-2} \right] \nonumber \\
& & + \left(\alpha^{(e,\nu)}\widehat{1} + \beta^{(e,\nu)}\widehat{n}\right)
\widehat{a}e^{i\varphi^{(e,\nu)}} + \widehat{a}^\dagger\left(\alpha^{(e,\nu)}
\widehat{1} + \beta^{(e,\nu)}\widehat{n}\right) e^{-i\varphi^{(e,\nu)}} \,,
\end{eqnarray}

\ni where

\begin{equation}
\widehat{\rho} = \frac{1}{\sqrt{29}}\left(\begin{array}{ccc} 1 & 0 & 0 \\ 0 & \sqrt{4}
& 0 \\ 0 & 0 & \sqrt{24} \end{array}\right)\;\;,\;\;\widehat{N} = \widehat{1} +
2 \widehat{n} = \left(\begin{array}{ccc} 1 & 0 & 0 \\ 0 & 3 & 0 \\ 0 & 0 & 5 
\end{array}\right)
\end{equation}

\ni and

\begin{equation}
\widehat{n} = \widehat{a}^\dagger\widehat{a} = \left(\begin{array}{ccc} 
0 & 0 & 0 \\ 0 & 1 & 0 \\ 0 & 0 & 2 \end{array}\right)\;\;,\;\;
\widehat{a} = \left(\begin{array}{ccc}  0 & 1 & 0 \\ 0 & 0 & \sqrt{2} \\
0 & 0 & 0 \end{array}\right)\;\;.
\end{equation}

\ni For charged leptons we will assume about the coupling constants $
\alpha^{(e)}/\mu^{(e)}$ and $\beta^{(e)}/\mu^{(e)}$ that the second term in
the matrix $\widehat{h}^{(e)}$ can be treated as a small perturbation of the 
first term. For neutrinos we will conjecture two alternative options: either 
({\it i}) the coupling constants $\alpha^{(\nu)}/\mu^{(\nu)}$ and $\beta^{(\nu
)}/\mu^{(\nu)}$ enable us to apply the perturbative treatment (similarly as for
charged leptons) and, in addition, $\varepsilon^{(\nu)\,2} \simeq 0 $, or ({\it
ii}) $\alpha^{(\nu)}/\mu^{(\nu)}$ only is a perturbative parameter and, 
additionally, $\varepsilon^{(\nu)\,2} \simeq 0 $.

 Note from Eqs. (21) that the "truncated" annihilation and creation matrices in
the family space, $\widehat{a}$ and $\widehat{a}^\dagger$, satisfy the familiar
commutation relations with $\widehat{n}$
 
\begin{equation}
\left[\widehat{a}\,,\;\widehat{n}\right] = \widehat{a}\;\;,\;\;
\left[\widehat{a}^\dagger\,,\;\widehat{n}\right] = -\widehat{a}^\dagger
\end{equation}

\ni and, additionally, the "truncation" identities

\begin{equation}
\widehat{a}^3 = 0\;\;,\;\; \widehat{a}^{\dagger\,3} = 0\;.
\end{equation}

\ni Thus, $\widehat{n}|n\rangle = n|n\rangle $ as well as $\widehat{a}|n
\rangle = \sqrt{n}|n - 1 \rangle$ and $\widehat{a}^\dagger|n \rangle = \sqrt{n 
+ 1}|n + 1\rangle\;\;(n = 0\,,\;1\,,\;2) $ , but  $\widehat{a}^\dagger|2\rangle
= 0 $ {\it i.e.}, $|3 \rangle = 0 $ (in addition to $\widehat{a}|0 \rangle = 0 
$ {\it i.e.}, $|-1 \rangle = 0 $). Evidently, $n = 0\,,\,1\,,\,2$ plays the 
role of an index $ i $ in our three--dimensional matrix calculations.

 For both labels $e$ and $\nu $ the mass matrix (18) can be written explicitly 
in the form

\begin{equation}
\widehat{M} = \frac{1}{29} \left(\begin{array}{ccc} 
\mu\varepsilon^2 & 2\alpha e^{i\varphi} & 0 \\ 2\alpha 
e^{-i\varphi} & 4\mu(80 + \varepsilon^2)/9 & 8(\alpha + \beta)\sqrt{3} 
e^{i\varphi} \\ 0 & 8(\alpha + \beta)\sqrt{3} e^{-i\varphi} & 24\mu(624 +
\varepsilon^2)/25 \end{array}\right)
\end{equation}

\ni (with obvious suppression of labels $ e $ and $\nu $).

 The unitary matrix $\widehat{U} \equiv \left(U_{i\,j}\right) $, diagonalizing 
the mass matrix $\widehat{M} \equiv \left(M_{i\,j}\right) $ according to the 
equation $\widehat{U}^{-1}\widehat{M}\widehat{U} = $ diag$(m_0\,,\;m_1\,,\;m_2)
$, has the form

\begin{equation}
\widehat{U} = \left(\begin{array}{ccc} A_0 & - A_1\frac{M_{01}}{M_{00} - m_1}
& A_2 \frac{M_{22} - m_2}{M_{21}}\frac{M_{01}}{M_{00} - m_2} \\ 
- A_0 \frac{M_{00} - m_0}{M_{01}} & A_1 & - A_2 \frac{M_{22} - m_2}{M_{21}} \\ 
A_0 \frac{M_{00} - m_0}{M_{01}}\frac{M_{21}}{M_{22} - m_0} & - A_1 
\frac{M_{21}}{M_{22} - m_1} & A_2 \end{array}\right)\; ,
\end{equation}

\ni where

\vfill\eject

\begin{eqnarray}
A_0 & = & \left\{1 + \frac{(M_{00} - m_{0})^2}{|M_{01}|^2}\left[ 1 + 
\frac{|M_{12}|^2}{(M_{22} - m_{0})^2}\right]\right\}^{-1/2}\; , \nonumber \\ 
A_1 & = & \left[1 + \frac{|M_{01}|^2}{(M_{00} - m_{1})^2} + 
\frac{|M_{12}|^2}{(M_{22} - m_{1})^2}\right]^{-1/2}\; , \nonumber \\ 
A_2 & = & \left\{1 + \frac{(M_{22} - m_{2})^2}{|M_{12}|^2}\left[ 1 + 
\frac{|M_{01}|^2}{(M_{00} - m_{2})^2}\right]\right\}^{-1/2}\; .
\end{eqnarray}

\ni The elements of lepton Cabibbo---Kobayashi---Maskawa matrix $\widehat{V}
\equiv (V_{i\,j}) = \widehat{U}^{(\nu)\,\dagger}\widehat{U}^{(e)}$ can be cal%
culated from the formulae $V_{i\,j} = \sum_k U^{(\nu)*}_{k\,i} {U}^{(e)}_{k\,j}
\;\;(i \,,\,j = 0\,,\,1\,,\,2$). Here, the secular equations det$(\widehat{M} - 
\widehat{1}\, m_i) = 0\;\;(i = 0\,,\,1\,,\,2) $ give

\begin{equation}
(M_{00} - m_i)(M_{11} - m_i)(M_{22} - m_i) = |M_{01}|^2\,(M_{22} - m_i) +
|M_{12}|^2\,(M_{00} - m_i) \;
\end{equation}

\ni due to $ M_{02} = 0 = M_{20}$. In particular, Eq. (27) implies that $M_{00}
- m_0 = 0 $ and $ (M_{11} - m_i)(M_{22} - m_i) = |M_{12}|^2\,\;(i = 1\,,\,2)$ if
$ M_{01} = 0 = M_{10}$. Also, $ (M_{00} - m_0)(M_{22} - m_0)^{-1}\rightarrow - 
|M_{01}|^2|M_{12}|^{-2} $ if $\mu \rightarrow 0 $.

\vspace{0.3cm}

\ni {\bf 3. Charged--lepton masses}

\vspace{0.3cm}

 Applying to the matrix $\widehat{M} $ given in Eq. (24) the first--order per%
turbative calculation with respect to its off--diagonal elements, we obtain

\begin{eqnarray}
m_0 & = & \frac{\mu}{29} \left[\varepsilon^2 - \frac{36}{320 - 5\varepsilon^2}
\left(\frac{\alpha}{\mu}\right)^2\right]\; , \nonumber \\ 
m_1 & = & \frac{\mu}{29} \left[\frac{4}{9}\left(80 + \varepsilon^2\right) +
\frac{36}{320 - 5\varepsilon^2}\left(\frac{\alpha}{\mu}\right)^2 - \frac{10800}
{31696 + 29\varepsilon^2}\left(\frac{\alpha+\beta}{\mu}\right)^2\right] \; ,
\nonumber \\ 
m_2 & = & \frac{\mu}{29} \left[\frac{24}{25}\left(624 + \varepsilon^2\right) +
\frac{10800}{31696 + 29\varepsilon^2}\left(\frac{\alpha+\beta}{\mu}
\right)^2\right] \;\;.
\end{eqnarray}

\ni These formulae imply the following mass sum rule: 

\begin{eqnarray}
m_2 & = & \frac{6}{125} \left(351 m_1 - 136 m_0 \right) \nonumber \\ 
& + & \frac{216}{3625} \left[ \frac{105300}{31696 + 29\varepsilon^2}\,\left(
\frac{\alpha + \beta}{\mu}\right)^2 - \frac{487}{320 - 5\varepsilon^2}\left(
\frac{\alpha}{\mu}\right)^2\right] \mu \;\;.
\end{eqnarray}

 In the case of charged leptons, the mass formulae (28) with $ m_e = m_0\,,\,
m_\mu = m_1\,,\, m_\tau = m_2 $ lead to

\vspace{-0.2cm}

\begin{eqnarray} 
m_\tau & = & 1776.80\;{\rm MeV} + O\left[\left(\alpha^{(e)}/\mu^{(e)}\right)^2
\right]\mu^{(e)} + O\left\{\left[\left(\alpha^{(e)} + \beta^{(e)}\right)/
\mu^{(e)}\right]^2 \right\} \mu^{(e)}\;,\nonumber \\
\varepsilon^{(e)\,2} & = & 0.172329 + O\left[\left(\alpha^{(e)}/\mu^{(e)}
\right)^2 \right] \;,\nonumber \\
\mu^{(e)} & = & 85.9924\;{\rm MeV} + O\left[\left(\alpha^{(e)}/\mu^{(e)}
\right)^2\right]\mu^{(e)} + O\left\{\left[(\alpha^{(e)} + \beta^{(e)})/
\mu^{(e)}\right]^2\right\}\mu^{(e)}\;,
\end{eqnarray}

\vspace{-0.2cm}

\ni if the experimental values of $ m_e$ and $ m_\mu$ [2] are used as an input.
Thus, the sum rule (29) gives

\vspace{-0.1cm}

\begin{equation}
m_\tau = \left[1776.80 + 9.20087\,\left(\frac{\alpha^{(e)}}{\mu^{(e)}}\right)^2
\right]\;{\rm MeV} \;,
\end{equation}

\vspace{-0.2cm}

\ni if we put for the sake of simplicity $\beta^{(e)} = 0$. With the experim\-%
ental value $m_\tau = 1777.00^{+0.030}_{-0.027} $ MeV, Eq. (31) shows that

\vspace{-0.1cm}
							
\begin{equation}
\left(\frac{\alpha^{(e)}}{\mu^{(e)}}\right)^2 = 0.022^{+0.033}_{-0.029} \;.
\end{equation}

\vspace{-0.1cm}

\ni So, as yet, the value of $\alpha^{(e)}$ is consistent with zero (of course,
from the viewpoint of our model, the acceptable lower error in Eq. (32) is 
$-0.022$).

 We can see that our model for $\widehat{M}^{(e)}$, even in the zero--order 
perturbative calculation, predicts excellently the mass $m_\tau$ [1].

\vspace{0.3cm}

\ni {\bf 4. Neutrino masses (the first option)}

\vspace{0.3cm}

 In the case of neutrinos consistent with our first option ($\alpha^{(\nu)}
/\mu^{(\nu)} $ and $\beta^{(\nu)}/\mu^{(\nu)} $ are perturbative parameters
and $\varepsilon^{(\nu)\,2} \simeq 0 $), the mass formulae (28) take the form

\begin{eqnarray} 
m_{\nu_e} & = & \frac{\mu^{(\nu)}}{29}\varepsilon^{(\nu)\,2} - \frac{9}{2320}
\left(\frac{\alpha^{(\nu)}}{\mu^{(\nu)}}\right)^2 \mu^{(\nu)} \simeq 0\;,
\nonumber \\
m_{\nu_\mu} & \simeq & \frac{320}{261} \mu^{(\nu)}\;, \nonumber \\
m_{\nu_\tau} & \simeq & \frac{14976}{725} \mu^{(\nu)}\;,
\end{eqnarray}

\ni if $\left(\alpha^{(\nu)}/\mu^{(\nu)}\right)^2 < 1$ and $\left(\beta^{(\nu)}
/\mu^{(\nu)}\right)^2 < 1 $. Here, the possible minus sign at $ m_{\nu_e} $ 
can be changed (if considered from the phenomenological point of view) into 
the plus sign since only $ m^2_{\nu_e} $ is relevant relativistically ({\it 
cf.} the Dirac equation). Note from Eqs. (33) that

\vspace{-0.1cm}

\begin{equation}
\frac{m_{\nu_\tau}}{m_{\nu_\mu}} \simeq \frac{2106}{125} = 16.8480\;, 
\end{equation}

\ni thus this ratio is practically equal to

\vspace{-0.2cm}

\begin{equation}
\frac{m_{\tau}}{m_{\mu}} \simeq \frac{1777.00}{105.658} = 16.8184\;.
\end{equation}

 The recent Super--Kamiokande experiments for atmospheric neutrinos [3] seem to
show that

\vspace{-0.1cm}

\begin{equation}
|m_{\nu_\tau}^2 - m_{\nu_\mu}^2| \sim (0.0003\;\;{\rm to}\;\;0.01)\;{\rm eV}^2
\;,
\end{equation}

\ni with the value $ 0.005\;{\rm eV}^2 $ being preferable (if mixing of $\nu^{
(m)}_\mu $and $\nu^{(m)}_\tau $ is maximal). In this case, Eqs.(33) give

\vspace{-0.1cm}

\begin{equation}
\mu^{(\nu)} \simeq \frac{\left( m_{\nu_\tau}^2 - m_{\nu_\mu}^2\right)^{1/2}}{
20.6201} \sim (0.0008\;\;{\rm to}\;\;0.005)\;{\rm eV}\;.
\end{equation}

\ni Then, from Eqs. (33) we predict

\vspace{-0.2cm}

\begin{eqnarray} 
m_{\nu_e} & \sim & (0.3\;\;{\rm to}\;\;2)\times 10^{-5}\left[9\varepsilon^{(
\nu)\,2} - \left(\frac{\alpha^{(\nu)}}{\mu^{(\nu)}}\right)^2\right]\;{\rm eV}
\simeq 0\;, \nonumber \\
m_{\nu_\mu} & \sim & (1\;\;{\rm to}\;\;6)\times 10^{-3}\;{\rm eV}\;, 
\nonumber \\
m_{\nu_\tau} & \sim & (0.2\;\;{\rm to}\;\;1)\times 10^{-1}\;{\rm eV}\;.
\end{eqnarray}

\ni Hence,

\vspace{-0.1cm}

\begin{equation}
m_{\nu_\mu}^2 - m_{\nu_e}^2 \simeq  m_{\nu_\mu}^2 \sim (0.1\;\;{\rm to}\;\;4)
\times 10^{-5}\;{\rm eV}\;.
\end{equation}

\ni Here, the sign "$\sim $" means approximate equality deduced with the use of
bounds (36).

 If we put tentatively

\begin{equation}
\left(\frac{\alpha^{(\nu)}}{\mu^{(\nu)}}\right)^2 \simeq \left(\frac{\alpha^{
(e)}}{\mu^{(e)}}\right)^2\;,
\end{equation}

\ni where the rhs is estimated as in Eq. (32), then from the first of Eqs. (38)
we obtain

\begin{equation}
|m_{\nu_e}| \sim (0.7^{+1}_{-0.9}\;\;{\rm to}\;\;4^{+6}_{-5})\times 10^{-7}
\;{\rm eV}
\end{equation}

\ni if $\varepsilon^{(\nu)\,2} = 0$ for the sake of simplicity (of course, from
the viewpoint of our model, the realistic lower errors in Eq. (41) are --0.7 
and --4,
rerspectively). In such a case, from Eqs. (33)

\begin{equation}
\frac{|m_{\nu_e}|}{m_{\nu_\mu}}\simeq 7^{+10}_{-9}\times 10^{-5}\;\;,
\end{equation}

\ni thus this ratio is much smaller than

\begin{equation}
\frac{m_{e}}{m_{\mu}} = \frac{0.510999}{105.658} = 4.83635\times 10^{-3}
\end{equation}

\ni (obviously, the realistic lower error in Eq. (42) is --7).

 We can see from Eqs. (38) that $m_{\nu_e} + m_{\nu_\mu} + m_{\nu_\tau} \simeq 
m_{\nu_\tau} \sim $ (0.02~ to~ 0.1) eV, so (if our model for $\widehat{M}^{(
\nu)}$ works) neutrinos cannot be candidates for hot dark matter, because such 
a possibility requires several eV for the neutrino mass sum [4].


\vspace{0.3cm}

\ni {\bf 5. Neutrino oscillations (the first option)}

\vspace{0.3cm}

 In order to calculate elements of the lepton Cabibbo---Kobayashi---Maskawa 
matrix $\widehat{V} \equiv \left(V_{i\,j}\right) = \widehat{U}^{(\nu)\,\dagger}
\widehat{U}^{(e)}$ we use the formulae $V_{i\,j} = \sum_k {U}^{(\nu)\,*}_{k
\,i}{U}^{(e)}_{k\,j}\;\;(i\,,\,j = 0\,,\,1\,,\,2) $, where the unitary matrices
$\widehat{U}^{(\nu,\,e)}$ are given by Eq. (25) in cooperation with Eq. (24) 
(here, the labels $\nu $ and $ e $ are made explicit). In $\widehat{M}^{(\nu)}
$ we put $\varepsilon^{(\nu)\,2} = 0 $, while in $\widehat{M}^{(e)}$ we have 
approximately $\varepsilon^{(e)\,2} \simeq 0 $. Then, in the lowest (linear) 
order in $\alpha^{(\nu)}/\mu^{(\nu)}$, $\beta^{(\nu)}/\mu^{(\nu)}$ and 
$\alpha^{(e)}/\mu^{(e)}$, $\beta^{(e)}/\mu^{(e)}$ we obtain


\begin{eqnarray}
V_{01} & = & -\frac{\sqrt{4}}{29}\left(\frac{\alpha^{(\nu)}}{m_{\nu_\mu}}
e^{i\varphi^{(\nu)}} - \frac{\alpha^{(e)}}{m_{\mu}} e^{i\varphi^{(e)}} \right)
= - V^*_{10}\;\,, \nonumber \\  
V_{12} & = & -\frac{\sqrt{192}}{29}\left(\frac{\alpha^{(\nu)}+\beta^{(\nu
)}}{m_{\nu_\tau}}e^{i\varphi^{(\nu)}} - \frac{\alpha^{(e)}+\beta^{(e)}}{m_{
\tau}}e^{i\varphi^{(e)}} \right) = - V^*_{21}\;\,, \nonumber \\  
V_{02} & = & 0 =  V_{20}\;\,, \nonumber \\  
V_{00} & = & V_{11} = V_{22} = 1 \;\,.
\end{eqnarray}


 Inserting the matrix elements (44) into Eq. (16), we get in the lowest 
(quadratic) order in $\alpha^{(\nu)}/\mu^{(\nu)}$, $\beta^{(\nu)}/\mu^{(\nu)}
$ and $\alpha^{(e)}/\mu^{(e)}$, $\beta^{(e)}/\mu^{(e)}$ the following 
neutrino--oscillation probabilities (in the vacuum):


\begin{eqnarray}
P\left(\nu_e \!\rightarrow \!\nu_\mu,t\right)\!\! & = & \!\!\frac{16}{841}
\left[\left( \frac{\alpha^{(\nu)}}{m_{\nu_\mu}}\right)^2 + \left(\frac{\alpha^{
(e)}}{m_{\mu}} \right)^2 -2 \frac{\alpha^{(\nu)}}{m_{\nu_\mu}} \frac{\alpha^{(
e)}}{m_{\mu}} \cos \left(\varphi^{(\nu)} - \varphi^{(e)}\right)\right]
\nonumber \\  
\!\!& &\!\! \times \sin^2 \left( \frac{m^2_{\nu_\mu} - m^2_{\nu_e}}{4|\vec{p}|}
\,t\right) \;\,, \nonumber \\  
P\left(\nu_\mu\! \rightarrow \!\nu_\tau,t \right)\!\! & = & \!\!
\frac{768}{841}\left[\!\left(\!\frac{\alpha^{(\nu)}\!\!+\!\beta^{(\nu)}}{m_{
\nu_\tau}}\right)^2\!\!\!+\left(\frac{\alpha^{(e)}\!\!+\!\beta^{(e)}}{m_{\tau}}
\right)^2\!\!\!-\!2\frac{\alpha^{(\nu)}\!\!+\!\beta^{(\nu)}}{m_{\nu_\tau}} 
\frac{\alpha^{(e)}\!\!+\!\beta^{(e)}}{m_{\tau}} \cos \left(\varphi^{(\nu)}\!\!
-\!\!\varphi^{(e)}\!\right)\!\right] \nonumber \\  
\!\!& & \!\!\times \sin^2 \left( \frac{m^2_{\nu_\tau} - m^2_{\nu_\mu}}{4|
\vec{p}|}\,t \right) \;\,, \nonumber \\  
P\left(\nu_e \rightarrow \nu_\tau,t \right)\!\! & = & \!\!0\;\,.
\end{eqnarray}


 If we make use of Eqs (28), neglecting there the terms~~$O\left[\left(
\alpha^{(e)}/\mu^{(e)}\right)^2\right] $ and also $O\left\{\left[\left(\alpha^{(e)} 
+ \beta^{(e)}\right)/ \mu^{(e)} \right]^2\right\} $~(what leads to the 
relations $ m_\mu \simeq (320/261)\mu^{(e)}$ and $ m_\tau \simeq (14976/725)
\mu^{(e)}$ up to terms proportional to $\varepsilon^{(e)\,2}\simeq 0\,$), we 
can conclude from Eqs. (45) that

\begin{eqnarray}
P\left(\nu_e \rightarrow \nu_\mu,t \right) & = & 
0.0126\left[\!\left( \frac{\alpha^{(\nu)}}{\mu^{(\nu)}}\right)^2 + \left( 
\frac{\alpha^{(e)}}{\mu^{(e)}} \right)^2 -2 \frac{\alpha^{(\nu)}}{\mu^{(\nu)}}
\frac{\alpha^{(e)}}{\mu^{(e)}} \cos \left(\varphi^{(\nu)} - \varphi^{(e)}\right)
\right] \nonumber \\  
& & \times \sin^2 \left( \frac{m^2_{\nu_\mu} - m^2_{\nu_e}}{4|\vec{p}|}
\,t\right) \;\,, \nonumber \\  
P\left(\nu_\mu \rightarrow \nu_\tau,t \right) & = & 
0.00214\,\left[\left(\frac{\alpha^{(\nu)}+\beta^{(\nu)}}{\mu^{(\nu)}}\right)^2
+ \left(\frac{\alpha^{(e)}+\beta^{(e)}}{\mu^{(e)}}\right)^2 \right. \nonumber \\
& & - 2 \left.\frac{\alpha^{(\nu)}+\beta^{(\nu)}}{\mu^{(\nu)}} 
\frac{\alpha^{(e)}+\beta^{(e)}}{\mu^{(e)}} \cos \left(\varphi^{(\nu)} - 
\varphi^{(e)}\right)\right] \nonumber \\ & & \times \sin^2 \left( 
\frac{m^2_{\nu_\tau} - m^2_{\nu_\mu}}{4|\vec{p}|}
\,t \right) \;\,. 
\end{eqnarray}

\ni Here, the factors $[\;\;]<1 $, so the order of amplitude of $P\left(\nu_\mu 
\rightarrow \nu_\tau,t \right) $ is smaller than $O(10^{-3})$.

 We can see that this result, valid in the case of our first option, appears
to be inconsistent with the experiments for atmospheric neutrinos 
[3,4,5] which seem to indicate that the order of amplitude of $P\left(\nu_\mu 
\rightarrow \nu_\tau,t \right) $ is $ O(1) $.

\vspace{0.3cm}

\ni {\bf 6. Neutrino masses (the second option)}

\vspace{0.3cm}

 In the case of neutrinos consistent with the second option (where $\alpha%
^{(\nu)}/\mu^{(\nu)}$ only is a perturbative parameter and $\varepsilon^{(\nu)
\,2} \simeq 0 $), the second term in the matrix $\widehat{h}^{(\nu)}$ given in
Eq. (19) cannot be treated as a small perturbation of the first term.

 When $\alpha^{(\nu)}/\mu^{(\nu)} = 0 $, the neutrino mass matrix (24) takes 
the unperturbed form

\begin{equation}
\widehat{M}^{(\nu)} = \frac{1}{29} \left(\begin{array}{ccc} 
\mu^{(\nu)}\varepsilon^{(\nu)\,2} & 0 & 0 \\ & & \\ 0 & 4\mu^{(\nu)}(80 + 
\varepsilon^{(\nu)\,2} )/9 & 8\beta^{(\nu)}\sqrt{3} e^{i\varphi} \\ & & \\ 0 & 
8\beta^{(\nu)}\sqrt{3} e^{-i\varphi} & 24\mu^{(\nu)}(624 + \varepsilon^{(\nu)\,
2})/25 \end{array}\right)\;\;.
\end{equation}

\ni Evidently, its eigenvalues can be found exactly, reading


\begin{eqnarray}
m_0\;\; & = & M^{(\nu)}_{00} = \frac{\mu^{(\nu)}}{29}\varepsilon^{(\nu)\,2} \;,
\nonumber \\ 
m_{1,2} & = & \frac{M_{11}^{(\nu)} + M_{22}^{(\nu)}}{2} \mp \left[
\left( \frac{M_{11}^{(\nu)} - M_{22}^{(\nu)}}{2}\right)^2 +|M_{12}^{(\nu)} |^2 
\right]^{1/2} \nonumber \\ 
& = & \left[ 10.9 \mp 0.478 \frac{\beta^{(\nu)}}{\mu^{(\nu)}} \sqrt{1 + \left(
20.3 \frac{\mu^{(\nu)}}{\beta^{(\nu)}} \right)^2 }\right] \mu^{(\nu)}\;,
\end{eqnarray}

\ni if $\varepsilon^{(\nu)\,2}<0.1\,$. These eigenvalues give three unperturbed
neutrino masses

\begin{equation}
m_{\nu_e} = m_0\;,\;\,m_{\nu_\mu} = m_1\;,\;\,m_{\nu_\tau} = m_2
\end{equation}

\ni if, by convention, we ascribe the minus sign in Eq. (48) to $ m_{\nu_\mu}$.
Note that in the limit of $\mu^{(\nu)} \rightarrow 0 $ Eqs. (48) give $ m_0 
\rightarrow 0 $ and $ m_{1,2} \rightarrow \mp |M^{(\nu)}_{12}| = \mp 0.478 
\beta^{(\nu)}$.

 From Eqs. (48) we can evaluate the difference of mass squared :

\begin{eqnarray}
m_2^2 - m_1^2 & = & 2 |M_{12}^{(\nu)}| \left( M_{11}^{(\nu)} + M_{22}^{(\nu)}
\right) \left[1 + \left( \frac{M_{11}^{(\nu)} - M_{22}^{(\nu)}}{2 
|M_{12}^{(\nu)} |}\right)^2 \right]^{1/2} \nonumber \\ 
& = &  20.9\, \beta^{(\nu)}\mu^{(\nu)} \sqrt{1 + \left(20.3 \frac{\mu^{(\nu)}}
{\beta^{(\nu)}} \right)^2 }\;,
\end{eqnarray}

\ni if $\varepsilon^{(\nu)\,2} < 0.1\, $. Note also from Eq. (48) that

\begin{equation}
\frac{M_{11}^{(\nu)} - m_1}{M_{12}^{(\nu)}} = X e^{-i\varphi}\;\;,\;\;
\frac{M_{22}^{(\nu)} - m_2}{M_{21}^{(\nu)}} = - X e^{i\varphi}\;\;,
\end{equation}

\ni where 

\vspace{-0.3cm}

\begin{eqnarray}
X & = & \frac{M_{11}^{(\nu)} - M_{22}^{(\nu)}}{2 |M_{12}^{(\nu)}|} + \left[1 + 
\left( \frac{M_{11}^{(\nu)} - M_{22}^{(\nu)}}{2 |M_{12}^{(\nu)} |}\right)^2
\right]^{1/2} \nonumber \\
& = & - 20.3 \frac{\mu^{(\nu)}}{\beta^{(\nu)}} + \left[1 + \left(20.3 
\frac{\mu^{(\nu)}}{\beta^{(\nu)}} \right)^2\right]^{1/2}\;,
\end{eqnarray}

\vspace{-0.1cm}

\ni if $\varepsilon^{(\nu)\,2} < 0.1\, $.  Here, $(M_{22}^{(\nu)} - m_i) 
M_{21}^{(\nu)\,-1} = M_{12}^{(\nu)} (M_{11}^{(\nu)} - m_i)^{-1}\;\;(i =1,2) $,
as it follows from the secular equations det$(\widehat{M}^{(\nu)} - \widehat{1}
m_i) = 0 $ with $ M^{(\nu)}_{02} = 0 = M^{(\nu)}_{20}$ and $M^{(\nu)}_{01} = 0 
= M^{(\nu)}_{10}$. Also $ M^{(\nu)}_{11} + M^{(\nu)}_{22} = m_1 + m_2 $ and $ 
M^{(\nu)}_{11} M^{(\nu)}_{22} - |M^{(\nu)}_{12}|^2 = m_1 m_2 $. Note that Eqs. 
(48) and (52) give $m_{1,2} = M^{(\nu)}_{11,22} \mp |M^{(\nu)}_{12}| X $.

 Now, let us assume that the neutrino mass matrix (47) is perturbed by the 
matrix

\vspace{-0.3cm}

\begin{equation}
\delta\widehat{M}^{(\nu)} = \left(\begin{array}{ccc} 0 & \delta M^{(\nu)}_{01} 
& 0 \\ \delta M^{(\nu)}_{10} & 0 & \delta M^{(\nu)}_{12} \\ 0 & \delta M^{(\nu)
}_{21} & 0 \end{array}\right) = \frac{\alpha^{(\nu)}}{29} \left(
\begin{array}{ccc} 0 & 2 e^{i\varphi^{(\nu)}} & 0 \\ 2 e^{-i\varphi^{(\nu)}} 
& 0 & 8 \sqrt{3} e^{i\varphi^{(\nu)}} \\ 0 & 8 \sqrt{3} e^{-i\varphi^{(\nu)}} 
& 0 \end{array}\right)\;\;.
\end{equation}

\vspace{-0.1cm}

\ni Remember that in the unperturbed mass matrix (47) $ M^{(\nu)}_{01} = 0 = 
M^{(\nu)}_{10}$, while $ M^{(\nu)}_{12} = (8\beta^{(\nu)} \sqrt{3}/29)\exp (i 
\varphi^{(\nu)}) = M^{(\nu)\,*}_{21}$. Then, the secular equations det$[\widehat
{M}^{(\nu)} +\delta \widehat{M}^{(\nu)} - \widehat{1}(m_i +\delta m_i)] = 0 $ 
give in the lowest (linear or quadratic) perturbative order in $\alpha^{(\nu)}/
\mu^{(\nu)}$ the following neutrino mass corrections:

\vspace{-0.2cm}

\begin{eqnarray}
\delta m_0\;\; & = &  - \frac{|\delta M^{(\nu)}_{01}|^2 M^{(\nu)}_{22}}{m_1 m_2}
\;,\nonumber \\ \delta m_{1,2} & = & \mp  \frac{|\delta M^{(\nu)}_{12}| | 
M^{(\nu)}_{12}|}{m_2 - m_1}\;,
\end{eqnarray}

\vspace{-0.1cm}

\ni where $|\delta M^{(\nu)}_{01}|^2 = 0.00476 \alpha^{(\nu)\,2}$, $ |\delta 
M^{(\nu)}_{12}| = 0.478 \alpha^{(\nu)}$, $ M^{(\nu)}_{22} = 20.7 \mu^{(\nu)}$  
and $ |M^{(\nu)}_{12}| = 0.478 \beta^{(\nu)}$. Here, we neglect all terms 
proportional to $ \varepsilon^{(\nu)\,2}$ (this is correct for $\varepsilon^{(
\nu)\,2} < 0.1 $). From Eqs. (54) it follows that $\delta m_0 + \delta m_1 + 
\delta m_2 = 0 $, as it should be because of tr$\,\delta \widehat{M}^{(\nu)} = 
0 $.  Note that in the limit of $\mu^{(\nu)}\rightarrow 0 $ Eqs. (54) give $
\delta m_0 \rightarrow 0 $ and $\delta m_{1,2} \rightarrow \mp |\delta M^{(\nu
)}_{12}|/2 $. Thus, in this limit $(m_1+\delta m_1)^2
= (m_2+\delta m_2)^2 $ as well as $ m^2_1 = m^2_2 $.

\vspace{0.3cm}

\ni {\bf 7. Neutrino oscillations (the second option)}

\vspace{0.3cm}

 The unitary matrix (25), diagonalizing the unperturbed neutrino mass matrix 
(47) according to the equation $\widehat{U}^{(\nu)\,-1}\widehat{M}^{(\nu)}
\widehat{U}^{(\nu)} = {\rm diag}(m_0\,,\,m_1\,,\,m_2)$, can be written as

\begin{equation}
\widehat{U}^{(\nu)} = \left(\begin{array}{ccc} 1 & 0 & 0 \\ 0 & A^{(\nu)}_1 & 
A^{(\nu)}_2 X e^{i\varphi^{(\nu)}} \\ 0 & -A^{(\nu)}_1 X e^{-i\varphi^{(\nu)}} 
& A^{(\nu)}_2 \end{array}\right)\;,
\end{equation}

\vspace{-0.2cm}

\ni where

\vspace{-0.2cm}

\begin{equation}
A^{(\nu)}_1 = \left(1 + X^2\right)^{-1/2} = A^{(\nu)}_2\;.
\end{equation}

\ni Here, X is given as in Eq. (51). Note that in the limit of $\mu^{(\nu)} 
\rightarrow 0 $ Eqs. (52) and (56) give $ X \rightarrow 1 $ and $ A^{(\nu)}_1 =
A^{(\nu)}_2 \rightarrow 1/\sqrt{2}$.

 Assuming tentatively that $\alpha^{(e)}$ and $\beta^{(e)}$, which are experim%
entally consistent with zero [{\it cf.} Eq. (32)], are really zero {\it i.e.}, 
$\widehat{U}^{(e)} = \widehat{1}$, we have

\vspace{-0.2cm}

\begin{equation}
\widehat{V} = \widehat{U}^{(\nu)\dagger} = \left( U^{(\nu)*}_{j\,i}\right)
\end{equation}

\ni with $\widehat{U}^{(\nu)} = \left( U^{(\nu)}_{i\,j}\right) $. Then, Eqs. 
(16) and (55) give the following unperturbed neutrino oscillation probabilities
(in the vacuum):

\vspace{-0.2cm}

\begin{eqnarray}
P\left(\nu_e \rightarrow \nu_\mu,t \right) & = & 0 = P\left(\nu_e \rightarrow 
\nu_\tau,t \right)\;,\nonumber \\ P\left(\nu_\mu \rightarrow \nu_\tau,t \right) 
& = & 4 \frac{X^2}{(1+X^2)^2} \sin^2 \left(\frac{m_{\nu_\tau}^2 - m_{\nu_\mu}^2}
{4|\vec{p}|}\,t\right)\;.
\end{eqnarray}


 Note from Eq. (52) that the oscillation amplitude $ 4 X^2 (1+X^2)^{-2}
\rightarrow 1 $ in the limit of $\mu^{(\nu)}/\beta^{(\nu)} \rightarrow 0 $ as 
then $ X \rightarrow 1 $. The atmospheric neutrino experiments seem to indicate
that this oscillation amplitude is of the order $ O(1) $, perhaps $\sim 1/2 $ 
[4]. So, taking $ 4 X^2 (1+X^2)^{-2} \sim 1/2 $ as an input, we estimate $ X 
\sim (3 - 2\sqrt2)^{1/2} = \sqrt2 - 1 $, what through Eq. (52) implies that

\vspace{-0.2cm}

\begin{equation}
20.3 \frac{\mu^{(\nu)}}{\beta^{(\nu)}} \sim 1
\end{equation}


\ni or $ \beta^{(\nu)}/\mu^{(\nu)} \sim 20.3 $ and $ \mu^{(\nu)}/\beta^{(\nu)} 
\sim 0.05\, $.

 Now, assuming as another input the Super--Kamiokande bound (36), we obtain from Eqs. 
(50) and (59)

\vspace{-0.2cm}
 
\begin{equation}
29.6 \mu^{(\nu)}\beta^{(\nu)} \sim \left(0.0003\;\;{\rm to}\;\; 0.01\right)
{\rm eV}^2\;.
\end{equation}

\vspace{-0.2cm}

\ni Of course, this relation excludes $\mu^{(\nu)} = 0 $, what would give $m^2_1
= m^2_2 $ as well as $(m_1 + \delta m_1)^2 = (m_2 + \delta m_2)^2$. Making 
use of Eqs. (59) and (60), we estimate

\vspace{-0.2cm}

\begin{equation}
\mu^{(\nu)} \sim \left(0.71\;\;{\rm to}\;\; 4.1\right)\times 10^{-3} {\rm eV}
\end{equation}

\vspace{-0.2cm}

\ni and

\vspace{-0.2cm}
 
\begin{equation}
\beta^{(\nu)} \sim \left(1.4\;\;{\rm to}\;\; 8.3\right)\times 10^{-2} {\rm eV}\;.
\end{equation}

 Finally, using the estimates (59) and (61), we can calculate from Eqs. (48) the
unperturbed neutrino masses

\vspace{-0.2cm}

\begin{eqnarray}
m_{0} & = & \frac{ \mu^{(\nu)}}{29}\varepsilon^{(\nu)\,2} \sim \left(0.24\;\;
{\rm to}\;\; 1.4\right)\times 10^{-4}\varepsilon^{(\nu)\,2} {\rm eV} \ll |m_1|
\;,\;\nonumber \\ 
m_{1} & \sim & -2.82 \mu^{(\nu)} \sim - \left(0.20\;\;{\rm to}\;\; 1.2\right)
\times 10^{-2} {\rm eV} \;,\;\nonumber \\ 
m_{2} & \sim & 24.6 \mu^{(\nu)} \sim \left(0.17\;\;{\rm to}\;\; 1.0\right)
\times 10^{-1} {\rm eV}\;,
\end{eqnarray}

\vspace{-0.1cm}

\ni if $\varepsilon^{(\nu)\,2} < 0.1\, $. The minus sign at $ m_1 $ is irrelevant
({\it cf.} the Dirac equation) and so, can be changed (if considered from the 
phenomenological point of wiew) into the plus sign.

 Similarly, from Eqs. (54) we can evaluate the neutrino mass corrections in 
terms of $\alpha^{(\nu)}/\mu^{(\nu)}$:

\begin{eqnarray}
\delta m_0\;\; & \sim & \,\;0.0014\left(\frac{\alpha^{(\nu)}}{\mu^{(\nu)}}\right)^2
\mu^{(\nu)} \sim \left(1.0\;\;{\rm to}\;\; 5.9\right)\times 10^{-6}
\left( \frac{\alpha^{(\nu)}}{\mu^{(\nu)}}\right)^2 {\rm eV} \;, \nonumber \\ 
\delta m_{1,2} & \sim & \mp\, 0.17 \frac{\alpha^{(\nu)}}{\mu^{(\nu)}}
\mu^{(\nu)} \sim \mp  \left(1.2\;\;{\rm to}\;\; 6.9\right)\times 10^{-4}\, 
\frac{\alpha^{(\nu)}}{\mu^{(\nu)}}\;{\rm eV} \;,
\end{eqnarray}

\ni if $\varepsilon^{(\nu)\,2} < 0.1\, $. Thus, $\delta m_0/m_0 \sim 4.2
\times 10^{-2} \left(1/\varepsilon^{(\nu)\,2}\right)\left(\alpha^{(\nu)}/
\mu^{(\nu)}\right)^2 $, $\delta m_1/m_1 \sim 6.0 \times10^{-2} \alpha^{(\nu)}/
\mu^{(\nu)}$ and $\delta m_2/m_2 \sim 6.9 \times 10^{-3} \alpha^{(\nu)}/\mu^{(
\nu)} $, what implies that on our accuracy level we get $ m_i + \delta m_i 
\sim m_i $ for $ i = 1\,,\;2$.

 We can see that the unperturbed result (58) for $ P\left(\nu_\mu \rightarrow 
\nu_\tau,t \right) $, valid in the case of our second option, is consistent 
with the experiments for atmospheric neutrinos [3,4,5], which suggest a 
large neutrino--oscillation amplitude of the order $ O(1) $. However, in the 
case of our second option, the vanishing $ P\left(\nu_e \rightarrow \nu_\mu,
t\right)$ and $ P\left(\nu_e \rightarrow \nu_\tau,t \right) $ raise a problem 
for solar neutrinos. Of course, the perturbed neutrino mass matrix $\widehat{M
}^{(\nu)} + \delta\widehat{M}^{(\nu)}$, as described by Eqs. (47) and (53), 
induces a perturbation $\delta\widehat{U}^{(\nu)}$ for the diagonalizing 
unitary matrix $\widehat{U}^{(\nu)}$ given in Eq. (55), and so, a perturbation 
$\delta \widehat{V}$ for the lepton Cabibbo---Kobayashi---Maskawa matrix $
\widehat{V} = \widehat{U}^{(\nu)\,\dagger}$. Obviously, when $\widehat{V} 
\rightarrow \widehat{V} + \delta\widehat{V}$ in consequence of $\widehat{M}^{(
\nu)} \rightarrow \widehat{M}^{(\nu)} + \delta \widehat{M}^{(\nu)}$, then

\vspace{-0.3cm}

\begin{eqnarray}
P\left(\nu_e \rightarrow \nu_\mu,t \right) & = & 0 \rightarrow \delta 
P\left(\nu_e \rightarrow \nu_\mu,t \right)\;,\nonumber \\ 
P\left(\nu_e \rightarrow \nu_\tau,t \right) & = & 0 \rightarrow \delta
P\left(\nu_e \rightarrow \nu_\tau,t \right)\;,\nonumber \\ 
P\left(\nu_\mu \rightarrow \nu_\tau,t \right) & \rightarrow & P\left(\nu_\mu 
\rightarrow \nu_\tau,t \right) + \delta P\left(\nu_\mu \rightarrow \nu_\tau,t 
\right) \;.
\end{eqnarray}

\ni If the realistic $\alpha^{(e)}$ and/or $\beta^{(e)}$ are not zero {\it i.%
e.}, $\widehat{U}^{(e)} \neq \widehat{1} $, then Eqs. (58) get also other 
corrections which will be discussed in detail in the next Section. The pertur%
bed $\widehat{V} +\delta\widehat{V}$, strengthened by the mechanism of neutrino
oscillations in the Sun matter [6,4,5], might help with the problem of solar 
neutrinos, practically not perturbing the oscillations (in the vacuum) of 
atmospheric neutrinos.

 The perturbation $\delta\widehat{V} = \left(\delta\widehat{U}^{(\nu)}\right)^{
\dagger}$ of the lepton Cabibbo---Kobayashi---Maskawa matrix $\widehat{V} = 
\widehat{U}^{(\nu)\,\dagger}$ (in the case of $\widehat{U}^{(e)} = \widehat{1}
$) can be calculated from Eq. (25) applied to the whole mass matrix $\widehat{M
}^{(\nu)} + \delta \widehat{M}^{(\nu)}$ given by Eqs. (47) and (53). Then, in 
this equation, $ M^{(\nu)}_{01} \rightarrow \delta M^{(\nu)}_{01}$ and $ M^{(
\nu)}_{12} \rightarrow M^{(\nu)}_{12} + \delta M^{(\nu)}_{12}$, and so, $A^{(
\nu)}_i \rightarrow A^{(\nu)}_i + \delta A^{(\nu)}_i $ as well as $ m_i 
\rightarrow m_i + \delta m_i\;\;(i = 0,1,2) $. Here, $\delta A^{(\nu)}_i $ are 
of the second order in $\alpha^{(\nu)}/\mu^{(\nu)} $, while $ \delta m_i $ are 
negligible. But, due to Eq. (54), the elements

\vspace{-0.2cm}

\begin{eqnarray}
U^{(\nu)}_{10} \rightarrow \delta U^{(\nu)}_{10} & = & - A^{(\nu)}_0 
\frac{M^{(\nu)}_{00} -m_0 - \delta m_0}{\delta M^{(\nu)}_{01}} = A^{(\nu)}_0
\frac{\delta m_0}{\delta M^{(\nu)}_{01}} \nonumber \\ & = & - A^{(\nu)}_0
\frac{|\delta M^{(\nu)}_{01}\!|\,M^{(\nu)}_{22}}{m_1 m_2}
\end{eqnarray}

\ni and $ U^{(\nu)}_{20} \rightarrow \delta U^{(\nu)}_{20}$ are of the first 
order in $\alpha^{(\nu)}/\mu^{(\nu)}$. Remember that $ A^{(\nu)}_0 = 1 $ and $
A^{(\nu)}_1 = A^{(\nu)}_2 = (1+X^2)^{-1/2}$. In this way, after some calcul%
ations, we obtain in the lowest (linear) perturbative order in $\alpha^{(\nu)}/
\mu^{(\nu)}$

\begin{equation}
\delta \widehat{U}^{(\nu)} = \left(\begin{array}{ccc} 
0 & A^{(\nu)}_1\frac{\delta M^{(\nu)}_{01}}{m_1} & A^{(\nu)}_2 \frac{\delta
M^{(\nu)}_{01} M^{(\nu)}_{12}}{m_2 (M^{(\nu)}_{22} - m_1)}\\ 
- \frac{\delta M^{(\nu)}_{10} M^{(\nu)}_{22}}{m_1 m_2 } & 0 & A^{(\nu)}_2 
\frac{\delta M^{(\nu)}_{12}}{M^{(\nu)}_{22} - m_1} \\ \frac{\delta M^{(\nu)}_{
10} M^{(\nu)}_{21}}{m_1 m_2 } & -A^{(\nu)}_1 \frac{\delta M^{(\nu)}_{21}}{M^{(
\nu)}_{22} - m_1} & 0 \end{array} \right) \;,
\end{equation}

\ni where $\delta M_{01}^{(\nu)} = (2\alpha^{(\nu)}/29)\exp(i\varphi^{(\nu)}) =
\delta M_{10}^{(\nu)\,*}$ and $\delta M^{(\nu)}_{12} = (8\alpha^{(\nu)}\sqrt{3}
/29)\exp (i \varphi^{(\nu)}) = \delta M^{(\nu)\,*}_{21}$, while $\widehat{U}^{(
\nu)}$ is given as in Eq. (55) with $ X\exp(i\varphi^{(\nu)}) = - \left( 
M^{(\nu)}_{22} - m_2\right)/M^{(\nu)}_{21}$. Here, all terms proportional to 
$\varepsilon^{(\nu)\,2}\simeq 0 $ are neglected (it is correct already for $
\varepsilon^{(\nu)\,2} < 0.1 $).

 The corrections $\delta P\left(\nu_e \rightarrow \nu_\mu,t \right)$, $\delta
P\left(\nu_e \rightarrow \nu_\tau,t \right)$ and $\delta P\left(\nu_\mu 
\rightarrow \nu_\tau,t \right) $ to the neutrino oscillation probabilities (in
the vacuum) can be evaluated from Eqs. (16) applied to the whole Cabibbo---%
Kobayashi---Maskawa matrix $\widehat{V} + \delta\widehat{V} = \widehat{U}^{(\nu
)\,\dagger} + \left(\delta \widehat{U}^{(\nu)}\right)^\dagger $ (in the case of
$\widehat{U}^{(e)} = \widehat{1}$) given by Eqs. (55) and (67). Then, after 
some calculations, we get in the lowest (linear or quadratic) perturbative 
order in $\alpha^{(\nu)}/\mu^{(\nu)}$ the following formulae:

\vspace{-0.1cm}

\begin{eqnarray}
\!\!\delta P\left(\nu_e \rightarrow \nu_\mu,t \right)\!\! & = & \!\!4|\delta 
V_{10}|^2 \left[\frac{M^{(\nu)}_{22}}{m_2}\sin^2 \left(\frac{m_{\nu_\mu}^2 - 
m_{\nu_e}^2}{4|\vec{p}|}\,t\right)\right. \nonumber \\ \!\!& + & \!\!\left.\!
\frac{|M^{(\nu)}_{12}|}{m_2} X \sin^2 \!\left(\frac{m_{\nu_\tau}^2\!-\!m_{\nu
_e}^2}{4|\vec{p}|}\,t\!\right)\! -\frac{|M^{(\nu)}_{12}\!|\,M^{(\nu)}_{22}}{m_2
^2}X \sin^2\!\left(\frac{m_{\nu_\tau}^2\! - \!m_{\nu_\mu}^2}{4|\vec{p}|}\,t
\right)\!\right]
\end{eqnarray}

\ni for the oscillations $\nu_e \rightarrow \nu_\mu $,

\vspace{-0.1cm}

\begin{eqnarray}
\delta P\left(\nu_e \rightarrow \nu_\tau,t \right)\!\! & = & \!\!4|\delta 
V_{10}|^2 \frac{|M^{(\nu)|}_{12}}{m_2} X \left[-\frac{|\,m_1|\,M^{(\nu)}_{22}}
{m_2 |M^{(\nu)}_{12}|} X \sin^2 \left(\frac{m_{\nu_\mu}^2\!- \!m_{\nu_e}^2}{4|
\vec{p}|}\,t\right)\right. \nonumber \\ \!\!& + & \!\!\left.\!\frac{|m_1|}{m_2}
\sin^2 \left(\frac{m_{\nu_\tau}^2\!-\!m_{\nu_e}^2}{4|\vec{p}|}\,t\!\right)\! + 
\!\frac{M^{(\nu)}_{22}}{m_2}\sin^2\!\left(\frac{m_{\nu_\tau}^2\!-\!m_{\nu_\mu}
^2}{4|\vec{p}|}\,t\right)\right] 
\end{eqnarray}

\ni for the oscillations $\nu_e \rightarrow \nu_\tau $, and

\vspace{-0.1cm}

\begin{eqnarray}
\!\!\delta P\left(\nu_\mu\!\!\rightarrow\!\!\nu_\tau,t \right) & = & 8|\delta V_{12}|
\frac{X}{(1+X^2)^{3/2}}\sin^2 \left(\frac{m_{\nu_\tau}^2 - m_{\nu_\mu}^2}{4|
\vec{p}|}\,t\right) \nonumber \\ & + & \!\!4|\delta V_{10}|^2
\frac{|m_1|}{m_2} \frac{X^2}{1+X^2}\left[\!\sin^2 \left(
\frac{m_{\nu_\mu}^2-m_{\nu_e}^2}{4|\vec{p}|}\,t\right)\! - \sin^2 
\left(\frac{m_{\nu_\tau}^2 - m_{\nu_e}^2}{4|\vec{p}|}\,t\right)\!\right]
\end{eqnarray}

\ni for the oscillations $\nu_\mu \rightarrow \nu_\tau $. These corrections 
are to be added to the unperturbed values (58). Here, the following numbers 
are involved:

\begin{eqnarray}
\!\!|\delta V_{10}|^2\!\! & = & \frac{|\delta M^{(\nu)}_{10}|^2}{m^2_1(1+X^2)}
\! = \!\frac{4}{841(1+X^2)} \left(\frac{\alpha^{(\nu)}}{m_1}\right)^2 \sim 5.25
\times 10^{-4}\left(\frac{\alpha^{(\nu)}}{\mu^{(\nu)}}\right)^2\,, \nonumber \\
\!\!\!|\delta V_{12}|\! & = & \frac{|\delta M^{(\nu)}_{12}|}{(M^{(\nu)}_{22}-
m_1)\sqrt{1+X^2}}\! =\!\frac{8\sqrt{3}}{29\sqrt{1+X^2}}\frac{\alpha^{(\nu)}}{
M^{(\nu)}_{22}-m_1}\! \sim \!1.88\times10^{-2}\frac{\alpha^{(\nu)}}
{\mu^{(\nu)}}
\end{eqnarray}

\ni as well as

\begin{eqnarray}
M^{(\nu)}_{11} = \frac{320}{261}\mu^{(\nu)} = 1.23\mu^{(\nu)} & , &
M^{(\nu)}_{22} = \frac{14976}{725} \mu^{(\nu)} = 20.7\mu^{(\nu)}\,,
\nonumber \\
|M^{(\nu)}_{12}| & = & \frac{8\sqrt{3}}{29} \beta^{(\nu)} \sim 9.70\mu^{(\nu)}
\end{eqnarray}

\ni and

\begin{equation}
m_1 \sim -2.78 \mu^{(\nu)}\;,\;m_2 \sim 24.7\mu^{(\nu)}\;,\;X \sim \sqrt{2} - 1
= 0.414\;.
\end{equation}

\ni The perturbative parameter $\alpha^{(\nu)}/\mu^{(\nu)}$ is free. In Eqs.
(71)---(73), the sign "$\sim $" denotes the estimate valid in the case of our 
input $ 4 X^2(1+X^2)^{-2} \sim 1/2 $ leading to the relation (59) for $
\beta^{(\nu)}/\mu^{(\nu)}$ {\it i.e.}, $\beta^{(\nu)}/\mu^{(\nu)} \sim 20.3\,$. Another
input is Eq. (60) giving for $\mu^{(\nu)}$ the value (61), $\mu^{(
\nu)} \sim (0.71\;\;{\rm to}\;\;4.1)\times 10^{-3}$ eV.

 We can see from Eqs. (68)---(70) and (71)---(73) that the corrections to the
neutrino--oscillation probabilities (58) (in the vacuum) are very small (for 
$\alpha^{(\nu)}/\mu^{(\nu)} < 1$). The largest of them is $\delta P\left(
\nu_\mu \rightarrow \nu_\tau,t \right) $. 

\vspace{0.3cm}

\ni {\bf 8. Conclusions and a proposal}

\vspace{0.3cm}

 In this paper, starting with the generic form (24) of lepton mass matrix, fol%
lowing from our texture dynamics expressed by Eqs. (18)---(21), we concentrated
mainly on neutrinos. For the parameters involved in this form we considered two
options: either ({\it i}) among the neutrinos $\nu_e\,,\,\nu_\mu\,,\,\nu_\tau $
practically only the neighbours mix and do it weakly, or ({\it ii}) practically 
only $\nu_\mu $ and $\nu_\tau $ mix and do it strongly. In both cases, we 
evaluated the neutrino masses, the lepton Cabibbo---Kobayashi---Maskawa matrix 
and the neutrino--oscillation probabilities (in the vacuum), expressing all 
these quantities in terms of few parameters determined essentially from the 
experimental data. In the second case, we calculated also the lowest--order 
perturbative corrections to these quantities, caused by possible weak mixing of
$\nu_e $ with $\nu_\mu $ and $\nu_\tau $. 

 The second option turned out to be consistent with the experiments for atmo\-%
spheric neutrinos [3,4,5] which seem to indicate a large $\nu_\mu \rightarrow 
\nu_\tau $ oscillation amplitude of the order $ O(1) $. Then, very small $\nu_e
\rightarrow \nu_\mu $ and $\nu_e \rightarrow \nu_\tau $ oscillation amplitudes 
were implied and so, apparently, must be much strenghtened in the Sun matter 
[6,4,5] in order to avoid the problem for solar neutrinos. In these calculatio%
ns, it was tentatively assumed that the nondiagonal part of the charged--lepton
mass--matrix, which is experimentally consistent with zero, is really zero.

 Let us add a remark bearing on the last question and concerning the values of
parameters involved in our generic form of mass matrix, when it is applied also
to the quarks $ u\,,\,c\,,\,t $ and $ d\,,\,s\,,\,b $. Such an application, as 
it was made in the second Ref. [1], led to the values

\vspace{-0.2cm}

\begin{equation}
\alpha^{(u)} \simeq 1740\; {\rm MeV}\;\;,\;\; \alpha^{(d)} + \beta^{(d)} \simeq 
405\; {\rm MeV}\;,
\end{equation}

\vspace{-0.1cm}

\ni when they were fitted to the experimental data for $ |V_{cb}| $ and $ |
V_{ub}/V_{cb}| $. If $\alpha^{(u)} : \alpha^{(d)} = |Q^{(u)}| : |Q^{(d)}| = 2$,
as was conjectured there, then

\vspace{-0.2cm}

\begin{equation}
\alpha^{(d)} \simeq 870 \;{\rm MeV}\;\;,\;\;\beta^{(d)} \simeq -465 \;{\rm MeV}
\;,
\end{equation}

\vspace{-0.1cm}

\ni what leaves $\beta^{(u)}$ unknown, unless also $\beta^{(u)} : \beta^{(d)} 
= 2 $ giving $\beta^{(u)} \simeq -930 $ MeV (at present, $\beta^{(u)}$ cannot 
be determined from the data directly). In the spirit of the relation $ 
\alpha^{(u)} : \alpha^{(d)} = |Q^{(u)}| : |Q^{(e)}|$ for quarks, the analogical
conjecture $\alpha^{(\nu)} : \alpha^{(e)} = |Q^{(\nu)}| : |Q^{(e)}| = 0 $ would
be natural for leptons, leaving now $\alpha^{(e)}$ as well as $\beta^{(\nu)}$ 
and $\beta^{(e)}$ free (to be determined from the neutrino and charged--lepton 
experiments).

 In Sections 6 and 7 we allowed for $\alpha^{(\nu)}$ to be different from zero,
but small ($\alpha^{(\nu)}/\mu^{(\nu)} < 1 $). Now, using partly the suggestion
that $\alpha^{(\nu)} : \alpha^{(e)} = 0 $, we might expect rather the inequal%
ity $0 \leq \alpha^{(\nu)} : \alpha^{(e)} \ll 1 $. If so, a new perturbation $
\delta\widehat{V}^{(e)}$ of the unperturbed $\widehat{V} = \widehat{U}^{(\nu)\,
\dagger}\widehat{U}^{(e)}$ with the trivial $\widehat{U}^{(e)} = \widehat{1}$ 
would arise, when $\widehat{U}^{(e)} \rightarrow \widehat{1} + \delta\widehat{U
}^{(e)}$ with a $\delta\widehat{U}^{(e)}$ proportional to $\alpha^{(e)}$ or 
$\alpha^{(e)} + \beta^{(e)}$. Such a new $\delta\widehat{V}^{(e)}$ should be 
more significant than our previous perturbation $\delta\widehat{V}^{(\nu)} = 
\left(\delta\widehat{U}^{(\nu)}\right)^\dagger$ proportional to $\alpha^{(\nu
)}$ (and discussed in detail in Section 7).

\vspace{-0.2cm}

 As our last item in this paper, let us evaluate the perturbation 

\begin{equation}
\delta\widehat{V}^{(e)} = \widehat{U}^{(\nu)\,\dagger}\delta\widehat{U}^{(e)}
\;,
\end{equation}

\vspace{-0.1cm}

\ni and also the related corrections to the neutrino--oscillation probabilities
(in the vacuum). Then,

\begin{equation}
\widehat{V} \rightarrow \widehat{V} + \delta \widehat{V}^{(e)} + \delta
\widehat{V}^{(\nu)} = \widehat{V} + \delta \widehat{V}^{(e)}
\end{equation}

\ni under the conjecture of comparatively negligible or even vanishing $
\delta \widehat{V}^{(\nu)}$.

 Since $\alpha^{(e)}/\mu^{(e)}$ and $\beta^{(e)}/\mu^{(e)}$ can be treated as 
perturbative parameters, we obtain from Eqs. (25), (24) and (28) in the lowest 
(linear) order in $\alpha^{(e)}/\mu^{(e)}$ and $(\alpha^{(e)} + \beta^{(e)})/
\mu^{(e)}$ the unitary matrix $\widehat{1}+ \delta\widehat{U}^{(e)}$ diagonal%
izing the mass matrix $\widehat{M}^{(e)} + \delta \widehat{M}^{(e)}$, where

\begin{equation}
\delta \widehat{U}^{(e)} = \frac{1}{29}\left(\begin{array}{ccc} 
0 &  2\frac{\alpha^{(e)}}{m_\mu}e^{i\varphi^{(e)}} & 0 \\ 
- 2\frac{\alpha^{(e)}}{m_\mu}e^{-i\varphi^{(e)}} & 0 & 8\sqrt{3}\frac{
(\alpha^{(e)}+\beta^{(e)})}{m_\tau}e^{i\varphi^{(e)}} \\ 0 & -8\sqrt{3}\frac{
(\alpha^{(e)}+\beta^{(e)})}{m_\tau}e^{-i\varphi^{(e)}} & 0 \end{array} \right) 
\end{equation}

\ni (what is consistent with Eqs. (44) if there $\alpha^{(\nu)} = 0 = \beta^{(
\nu)}$ formally). Here, $\widehat{M}^{(e)} = $ diag$(m_e\,,\,m_\mu\,,\,m_\tau)$
contains the unperturbed charged--lepton masses

\begin{equation}
m_e = \frac{\mu^{(e)}}{29}\varepsilon^{(e)\,2}\;,\;m_\mu = \frac{\mu^{(e)}}{29}
\frac{4}{9}\left(80 +\varepsilon^{(e)\,2}\right)\;,\;m_\tau = \frac{\mu^{(e)}
}{29}\frac{24}{25}\left(624 +\varepsilon^{(e)\,2}\right)
\end{equation}

\ni [undistiguished, as yet, from their experimental values, as seen from 
Eqs. (31) and (32)], whereas

\begin{equation}
\delta \widehat{M}^{(e)} = \frac{1}{29}\left(\begin{array}{ccc} 
0 &  2\alpha^{(e)} e^{i\varphi^{(e)}} & 0 \\ 2\alpha^{(e)} e^{-i\varphi^{(e)}} 
& 0 & 8\sqrt{3}(\alpha^{(e)}+\beta^{(e)}) e^{i\varphi^{(e)}} \\ 0 & 8\sqrt{3}
(\alpha^{(e)}+\beta^{(e)}) e^{-i\varphi^{(e)}} & 0 \end{array} \right) 
\end{equation}

\ni is the perturbation. Note that in the lowest (quadratic) perturbative order
the perturbed masses $m_e + \delta m_e\,,\,m_\mu + \delta m_\mu\,,\,m_\tau
+ \delta m_\tau$ are given as in Eqs. (28).

 In the next step we make use of Eqs. (55) and (78) to calculate $\delta
\widehat{V}^{(e)} \equiv (\delta V_{i\,j}^{(e)}) = \widehat{U}^{(\nu)\,
\dagger} \delta \widehat{U}^{(e)}$. The result is

\begin{eqnarray}
\delta V^{(e)}_{01} & = & \frac{2}{29}\frac{\alpha^{(e)}}
{m_\mu}e^{i\varphi^{(e)}}\;,\;\delta V^{(e)}_{10} = -\frac{2}{29\sqrt{1+X^2}}
\frac{\alpha^{(e)}}{m_\mu}e^{-i\varphi^{(e)}}\;, \nonumber \\ \delta 
V^{(e)}_{12} & = & \frac{8\sqrt{3}}{29\sqrt{1+X^2}}
\frac{\alpha^{(e)}+\beta^{(e)}}{m_\tau}e^{i\varphi^{(e)}} = -\delta 
V^{(e)\,*}_{21}\;, \nonumber \\ \delta V^{(e)}_{02} & = & 0\;,\; \delta 
V^{(e)}_{20} = -\frac{2X}{29\sqrt{1+X^2}}\frac{\alpha^{(e)}}{m_\mu}
e^{-i(\varphi^{(\nu)}+\varphi^{(e)})}\;, \nonumber \\ \delta V^{(e)}_{00} 
& = & 0\;,\;\delta V^{(e)}_{11} = \frac{8\sqrt{3}X}{29\sqrt{1+X^2}}
\frac{\alpha^{(e)}+\beta^{(e)}}{m_\tau}e^{i(\varphi^{(\nu)}-\varphi^{(e)})} 
= \delta V^{(e)\,*}_{22}\;.
\end{eqnarray}

 Finally, we apply Eqs. (16) to the whole lepton Cabibbo---Kobayashi---Maskawa 
matrix $\widehat{V} + \delta \widehat{V}^{(e)}$, where $\widehat{V} = \widehat{
U}^{(\nu)\,\dagger}$ and $\delta \widehat{V}^{(e)}$ are given by Eqs. (55) and
(81), respectively (and $\delta\widehat{V}^{(\nu)} = \left(\delta\widehat{U}^{(
\nu)}\right)^\dagger$ is neglected). Then, after some calculations, we obtain 
in the lowest (linear or quadratic) perturbative order in $\alpha^{(e)}/\mu^{(
e)}$ and $(\alpha^{(e)}+\beta^{(e)})/\mu^{(e)}$ the following corrections to 
the neutrino--oscillation probabilities (in the vacuum):

\begin{equation}
\delta P\left(\nu_e \rightarrow \nu_\mu,t \right) = \frac{16}{841(1+X^2)} 
\left(\frac{\alpha^{(e)}}{m_\mu}\right)^2 \sin^2 \left(\frac{m_{\nu_\mu}^2 - 
m_{\nu_e}^2}{4|\vec{p}|}\,t\right) 
\end{equation}

\ni for the oscillations $\nu_e \rightarrow \nu_\mu$,

\begin{equation}
\delta P\left(\nu_e \rightarrow \nu_\tau,t \right) = \frac{16 X^2}{841(1+X^2)} 
\left(\frac{\alpha^{(e)}}{m_\mu}\right)^2 \cos^2 \left(\frac{m_{\nu_\mu}^2 - 
m_{\nu_e}^2}{4|\vec{p}|}\,t\right) 
\end{equation}

\ni for the oscillations $\nu_e \rightarrow \nu_\tau$, and

\begin{eqnarray}
\delta P\left(\nu_\mu \rightarrow \nu_\tau,t \right) & = & \frac{64\sqrt{3} X}
{29(1+X^2)}\frac{\alpha^{(e)}+\beta^{(e)}}{m_\tau}\cos\left(\varphi^{(\nu)}-
\varphi^{(e)}\right) \sin^2 \left(\frac{m_{\nu_\tau}^2 - m_{\nu_\mu}^2}{4|
\vec{p}|}\,t\right) \nonumber \\ & + & \frac{16 X^2}{841(1+X^2)}\left(
\frac{\alpha^{(e)}}{m_\mu}\right)^2 \left[ \sin^2\left(\frac{m_{\nu_\mu}^2 - 
m_{\nu_e}^2}{4|\vec{p}|}\,t\right) - \sin^2\left(\frac{m_{\nu_\tau}^2 - 
m_{\nu_e}^2}{4|\vec{p}|}\,t\right)\right] \nonumber \\ & &
\end{eqnarray}

\ni for the oscillations $\nu_\mu \rightarrow \nu_\tau$. Note that the same 
time--argument appears in Eqs. (82) and (83). Also notice the presence of 
unknown phase factor $\cos(\varphi^{(\nu)}-\varphi^{(e)})$ in Eq. (84) that 
becomes 1 if $\varphi^{(\nu)} = \varphi^{(e)}$ as {\it e.g.} for

\begin{equation}
\varphi^{(\nu)} = \varphi^{(e)} = 0\;.
\end{equation}

\ni Of course, these corrections are to be added to the unperturbed values 
(58). In Eqs. (82), (83) and (84) there appear the numerical coefficients

\begin{eqnarray}
\frac{16}{841(1+X^2)} \left(\frac{\alpha^{(e)}}{m_\mu}\right)^2 \sim 2.4\times 
10^{-4} & , & \frac{16 X^2}{841(1+X^2)}\left(\frac{\alpha^{(e)}}{m_\mu}\right)^2
\sim 4.1\times 10^{-5}\,, \nonumber\\
\;\frac{64\sqrt{3} X}{29(1+X^2)}\frac{\alpha^{(e)}+\beta^{(e)}}{m_\tau} & \sim
& 9.7\times 10^{-3}\;.
\end{eqnarray}

\ni To evaluate these coefficients we put $\beta^{(e)} = 0 $ for the sake of 
simplicity, and then took the central value $(\alpha^{(e)}/\mu^{(e)})^2 = 0.022 
$ deduced in Eq. (32) from the experimental value of $ m_\tau $. We used also 
our input $ 4 X^2 (1 + X^2)^{-2} \sim 1/2 $ implying $ X \sim \sqrt{2} - 1 $. When
the experimental errors in Eq. (32) are taken into account, the first coeffic%
ient (86) becomes $ 2.4^{+3.6}_{-2.4}\times 10^{-4} $.

We can conclude from Eq. (82) that the predicted oscillations $\nu_e\rightarrow
\nu_\mu $ (in the vacuum) are very small, and similar in magnitude to those 
derived in the case of our first option [{\it cf.} Eqs. (45)].Thus, the 
effect of neutrino oscillations in the Sun matter still appears to be needed. 
Evidently, the oscillations $\nu_e \rightarrow \nu_\mu $ caused by $ \delta 
\widehat{V}^{(\nu)} = \left(\delta \widehat{U}^{(\nu)} \right)^\dagger $ 
are comparatively negligible or even vanish if $ 0\leq\alpha^{(\nu)}\ll
\alpha^{(e)}$ [{\it cf.} Eq. (68)]. Further, from Eq. (84) it follows that the 
predicted correction to the overwhelming unperturbed oscillations $\nu_\mu 
\rightarrow \nu_\tau $ [{\it cf.} Eq. (58)] is larger in magnitude than 
the oscillations $\nu_e \rightarrow \nu_\mu $, and also larger than the 
oscillations $\nu_\mu \rightarrow \nu_\tau $ obtained in the case of our 
first option [{\it cf.} Eq. (45)]. Again, the correction caused by $\delta 
\widehat{V}^{(\nu)} = \left(\delta \widehat{U}^{(\nu)}\right)^\dagger$ is 
comparatively negligible or even vanishes [{\it cf.} Eq. (70)].

 Thus, the atmospheric neutrino experiments, if interpreted in terms of our 
"texture dynamics", seem to transmit an important message about strong mixing 
of $\nu_\mu $ and $\nu_\tau $ neutrinos and, on the other hand, their weak 
mixing with $\nu_e $. However, such a strong mixing cannot be really maximal as
then the degeneration $ m_{\nu_\mu}^2 = m_{\nu_\tau}^2 $ appears, excluding the
experimentally suggested large oscillations $\nu_\mu \rightarrow \nu_\tau $. 
{\it A priori}, some small oscillations  $\nu_e \rightarrow \nu_\mu $ (in the 
vacuum) may be caused by both factor matrices $\widehat{U}^{(\nu)\,\dagger}$ 
and $\widehat{U}^{(e)}$ in the lepton Cabibbo---Kobayashi---Maskawa matrix $
\widehat{V}$. In this Section of our paper we conjectured that $\widehat{U}^{(e
)}$ is practically responsible for such small oscillations (in the vacuum).

\vfill\eject

~~~~
\vspace{0.5cm}

{\bf References}

\vspace{1.0cm}

{\everypar={\hangindent=0.5truecm}
\parindent=0pt\frenchspacing

1.~W.~Kr\'{o}likowski, {\it Acta Phys. Pol.}, {\bf B 27}, 2121 (1996);
{\bf B 28}, 1643 (1997); and references therein. 

2.~{\it Review of Particle Properties}, {\it Phys.Rev.} {\bf D 54}, 1 (1996), 
Part I. 

3.~D. Kie{\l}czewska, {\it private communication}.

4.~For a review {\it cf.} G. Gelmini, E. Roulet, {\it Neutrino Masses}, 
CERN--TH 7541/94 (December 1994).

5.~Y. Suzuki, {\it Neutrino Masses and Oscillations~~---~~Experiment}, in {\it
Proc. of the 28th Inter. Conf. on High Energy Physics, Warsaw 1996}, eds. 
Z.~Ajduk and A.K.~Wr\'{o}blew\-ski, World Scientific Pub. 1997; A.Y.~Smirnow, 
{\it Neutrino Masses and Oscillations --- Theory}, in the same {\it Proceedings
}.

6.~S.P.~Mikheyev and A.Y.~Smirnow, {\it Sov. Jour. Nucl. Phys.}, {\bf 42}, 913
(1985); L.~Wolfenstein, {\it Phys.Rev.} {\bf D 17}, 2369 (1978).
\vfill\eject

\end{document}